\DocumentMetadata{
	lang		= eng-US, 	
	pdfstandard	= A-3u,		
	pdfversion	= 1.7, 		
}

\documentclass[colorlinks,upint,subscriptcorrection,varvw,hyphenate,balance,greek,russian,vietnamese,german]{asmeconf} 

\hypersetup{%
	pdfauthor={Michael Acquah, Zheng Liu},									  
	pdftitle={Cooling Channel Design Optimization for High Power Multi-Chip Packages},                  
	pdfkeywords={Thermal management, microchannel cooling, cooling channel design, optimization, multi-chip systems},
	pdfsubject = {asmeconf LaTeX},			  
}

\allowdisplaybreaks 

\begin{document}


\ConfName{Proceedings of the ASME 2026\linebreak International Design Engineering Technical Conferences and Computers and Information in Engineering Conference}
\ConfAcronym{IDETC/CIE2026}
\ConfDate{August 23--26, 2026} 
\ConfCity{Houston, TX}
\PaperNo{DETC2026-195024}

%

\title{Cooling Channel Design Optimization for High Power Multi-chip Packages} 
 
%
%
%

\SetAuthors{%
  Michael Acquah\affil{1}, Zheng Liu\affil{2}\CorrespondingAuthor{zhengtl@umich.edu}
}
\SetAffiliation{1}{Department of Mathematics and Statistics,
  University of Michigan-Dearborn, Dearborn, Michigan, USA}
\SetAffiliation{2}{Department of Industrial and Manufacturing Systems Engineering,
  University of Michigan-Dearborn, Dearborn, Michigan, USA}



\maketitle





\begin{abstract}
Thermal management is a major challenge in next-generation high-performance computing systems, particularly for heterogeneous multi-chip packages such as the NVIDIA GB200 Grace Blackwell Superchip. In this work, a physics-based computational framework is developed to optimize embedded cooling channel layouts for high-power multi-chip modules. The model couples steady-state heat conduction with a porous media-based representation of coolant transport, coupled with a row-wise coolant energy balance, to estimate chip temperature fields within microchannel networks. Unlike conventional designs, an interdigitated cooling architecture is parameterized using geometric variables, including channel count, width, and expansion over chip regions, enabling systematic design exploration. To enable efficient optimization, a surrogate-based approach is employed to approximate the relationship between geometric parameters and temperature metrics. The resulting model is optimized using a mixed-integer quadratic programming algorithm to minimize a weighted objective based on peak and average chip temperatures. To improve physical relevance, channel placement is further constrained to increase cooling coverage near GPU regions, where thermal loads are highest. The framework is applied to a representative multi-chip configuration based on NVIDIA GB200 architecture, consisting of two graphics processing units and one central processing unit. The results demonstrate that the optimal design reduces the peak chip temperature by 140.45$^\circ$C and the average chip temperature by 35.87$^\circ$C compared to the baseline configuration. The optimized symmetric architecture achieves an absolute maximum temperature of 79.91$^\circ$C and an average temperature of 66.27$^\circ$C, both of which remain below the recommended 95$^\circ$C operating limit for high-performance processors. This work provides a practical and extensible approach for topology-aware cooling design in advanced electronic packaging systems.
\end{abstract}


\begin{nomenclature}

\EntryHeading{Roman letters}

\entry{$T$}{Temperature [$^\circ$C]}
\entry{$T_{max}$}{Maximum temperature [$^\circ$C]}
\entry{$T_{avg}$}{Average temperature [$^\circ$C]}
\entry{$k$}{Thermal conductivity [W m$^{-1}$ K$^{-1}$]}
\entry{$q$}{Heat flux [W m$^{-2}$]}
\entry{$h$}{Convective heat transfer coefficient [W m$^{-2}$ K$^{-1}$]}
\entry{$\dot{m}$}{Total coolant mass flow rate [kg s$^{-1}$]}
\entry{$c_p$}{Specific heat capacity [J kg$^{-1}$ K$^{-1}$]}
\entry{$J$}{Objective function}
\entry{$N$}{Number of channels}
\entry{$W$}{Channel width [m]}
\entry{$E$}{Channel expansion width over chips [m]}
\entry{$L_x$}{Domain length in the $x$ direction [m]}
\entry{$L_y$}{Domain length in the $y$ direction [m]}
\entry{$\mathbf{v}$}{Coolant velocity vector [m s$^{-1}$]}
\entry{$|\mathbf{v}|$}{Coolant velocity magnitude [m s$^{-1}$]}
\entry{$K$}{Permeability of porous medium [m$^2$]}

\EntryHeading{Greek letters}

\entry{$\nabla$}{Gradient operator}
\entry{$\nabla^2$}{Laplacian operator}
\entry{$\mu$}{Dynamic viscosity [Pa s]}
\entry{$\rho$}{Coolant density [kg m$^{-3}$]}

\EntryHeading{Subscripts}

\entry{$amb$}{Ambient condition}
\entry{$cpu$}{Central processing unit}
\entry{$gpu$}{Graphics processing unit}
\entry{$in$}{Inlet}
\entry{$out$}{Outlet}

\end{nomenclature}


\section{Introduction}

The rapid increase in computational power demanded by modern processors, particularly graphics processing units (GPUs) and artificial intelligence accelerators, has led to significant growth in chip power density \cite{tuckerman2005high,liang2020high}. As transistor counts and switching frequencies continue to rise, efficient thermal management has become a critical challenge in high-performance computing systems \cite{mudawar2002assessment}. Excessive temperatures not only degrade device performance but can also reduce reliability and shorten the operational lifetime of electronic components \cite{tuckerman2005high, zhu2024machine}. Consequently, the design of effective cooling strategies has become an important aspect of modern chip and package design.

Traditional cooling methods, such as air cooling and heat sinks, are increasingly insufficient for handling the high heat fluxes generated by contemporary processors \cite{mudawar2002assessment,natarajan2009thermal}. Liquid cooling technologies, particularly microchannel-based cooling systems, have therefore gained considerable attention due to their ability to remove large heat loads within compact footprints \cite{tuckerman2005high,kandlikar2005high,thome2004boiling}. In microchannel cooling, coolant is routed through small channels embedded within or attached to the chip substrate, allowing heat to be transported away from localized high-power regions \cite{tuckerman2005high,colgan2007practical}. Comprehensive reviews of heat transfer and fluid flow in microchannels are available in Sobhan and Garimella and related surveys \cite{sobhan2001comparative, mudawar2002assessment}.

A major design challenge in microchannel cooling systems is the distribution of coolant flow across multiple heat sources. In multi-chip systems containing several high-power components, such as central processing units (CPUs) and GPUs, poorly distributed coolant flow can lead to thermal non-uniformity and the formation of localized hotspot \cite{liang2020high,peles2005forced}. Various channel configurations have been explored to address this issue, including parallel microchannels, tree-like networks, and manifold-based distribution structures \cite{escher2009efficiency,zhao2002analysis}. Among these approaches, manifold-style channel layouts have attracted interest because they can provide more balanced coolant delivery to multiple heat-generating regions \cite{escher2009efficiency,zhao2002analysis}.
Early work on microchannel heat exchanger optimization established that channel geometry directly governs the tradeoff between thermal resistance and pumping power, providing a foundation for subsequent design studies \cite{harpole1991micro}. Parametric and optimization studies of microchannel heat sink geometries have demonstrated that channel dimensions and layout significantly affect thermal performance, motivating the use of systematic design frameworks for channel layout selection \cite{wei2003optimization}.

Many computational studies of chip cooling rely on simplified thermal models that primarily consider heat diffusion within the substrate \cite{zhao2002analysis,kong2023additively}. While such models are useful for initial analysis, they do not explicitly represent coolant transport within the channel network. As a result, these approaches may not fully capture the interaction between thermal transport and coolant distribution in liquid-cooled systems, particularly in geometries where coolant connectivity directly affects hot-spot formation \cite{zhao2002analysis,escher2009efficiency,colgan2007practical, chen2025additive}.

In this work, we present a physics-based computational framework that couples steady-state heat conduction with a reduced-order representation of coolant transport and a row-wise coolant energy balance to evaluate temperature fields within explicitly defined channel networks. The framework is applied to a multi-chip package representative of modern high-power AI systems, specifically inspired by the NVIDIA GB200 Grace Blackwell Superchip, which integrates multiple high-performance processors within a compact footprint. 

Unlike conventional studies that focus on fixed channel layouts, this work formulates the cooling design problem as an optimization problem. An interdigitated channel architecture is parameterized using key geometric variables, including channel count, channel width, and expansion width over chip regions. A surrogate-based optimization approach is employed, where temperature responses are approximated as functions of these design variables and optimized using the Mixed-Integer Quadratic Programming (MIQP) algorithm to minimize a weighted objective function based on peak and average chip temperatures. To improve physical relevance, channel placement is further constrained to increase cooling coverage near GPU regions, where thermal loads are highest.

The primary contributions of this work are: (1) developed a coupled thermal model coolant transport representation for multi-chip systems, and (2) established an optimization-driven channel design framework that identifies cooling layouts capable of significantly reducing peak and average chip temperatures under realistic operating conditions.

\section{Problem Description}

This study investigates the thermal management of a multi-chip electronic package representative of modern high-power systems, such as the NVIDIA GB200 Grace Blackwell Superchip. The package consists of two GPUs and one CPU distributed over a square substrate of dimensions $12 \times 120~\text{mm}$. The system is modeled as a two-dimensional domain, which is appropriate when in-plane thermal conduction dominates, and the substrate thickness is small relative to its lateral dimensions \cite{zhao2002analysis,van2020embedded,liu2026generative}. The computational domain is discretized using a uniform $200 \times 200$ grid.

\begin{figure}[htbp]
\centering
\includegraphics[width=0.95\linewidth]{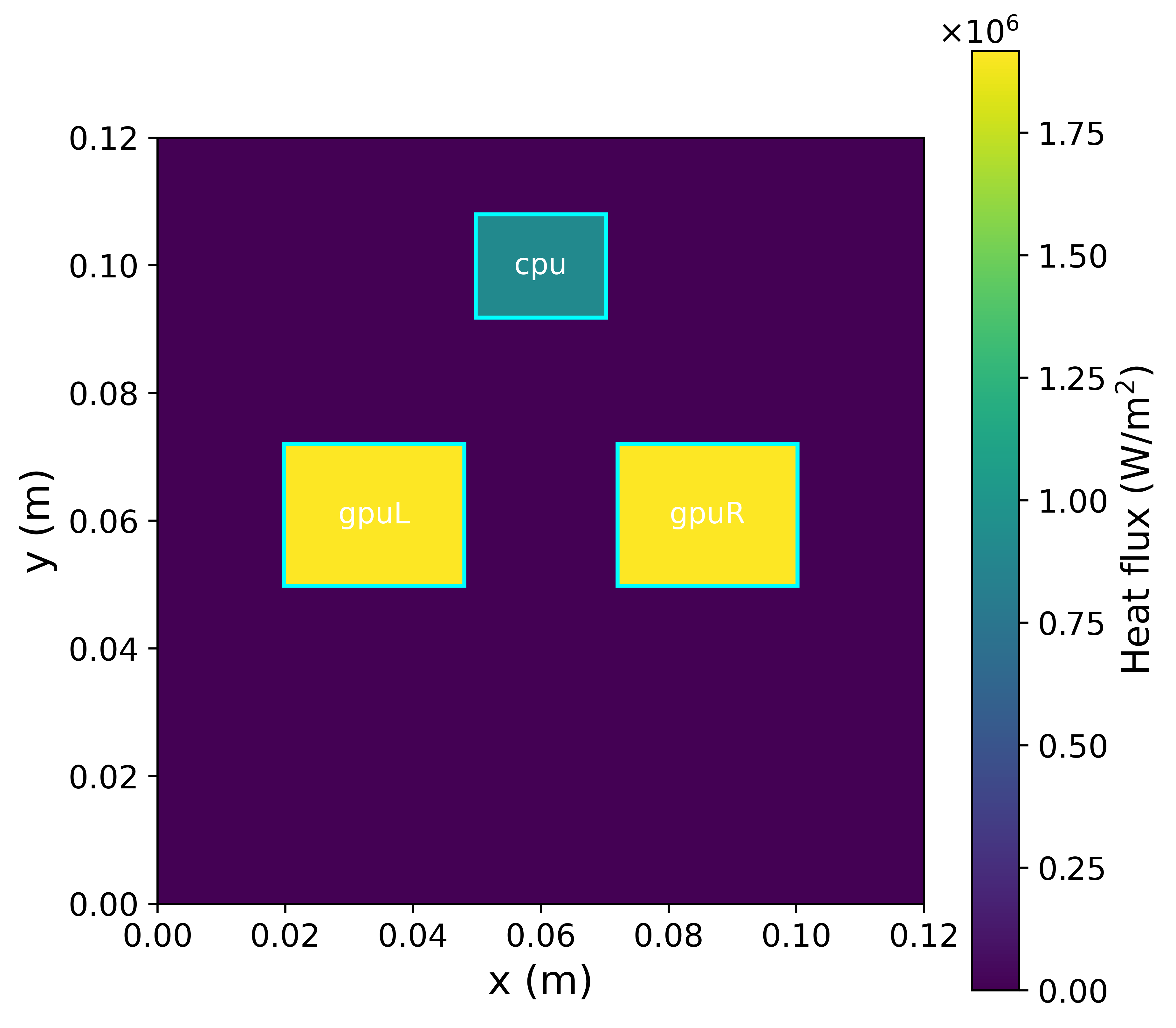}
\caption{Power density distribution within the computational domain showing the spatial arrangement of two GPUs and one CPU.}
\label{fig:layout}
\end{figure}

The component layout is shown in Fig.~\ref{fig:layout}. Each GPU dissipates $1200~\text{W}$, while the CPU dissipates $300~\text{W}$ \cite{fibermallDeepDive}. The GPUs are positioned symmetrically within the domain, while the CPU is located toward the upper region. Within each chip footprint, heat generation is modeled as a uniform surface heat flux applied across the corresponding area \cite{song2022case}.

Cooling is provided by a network of microchannels embedded within the substrate. Coolant enters the domain through a $25.4~\text{mm}$ wide inlet port located at the bottom center and exits through a corresponding outlet port at the top center. Within channel regions, heat removal is modeled using an effective convective heat transfer coefficient of $h = 1.0 \times 10^{5}~\text{W m}^{-2}\text{K}^{-1}$, representing aggressive liquid cooling conditions typical of microchannel heat sinks \cite{tuckerman2005high, escher2009efficiency, van2020embedded}. Outside the channels, heat transfer occurs through in-plane conduction within the substrate.

\section{Methodology}

The thermal framework developed for the analysis and optimization of embedded cooling-channel layouts in a high-power multi-chip package, integrating steady-state in-plane heat conduction, effective convective heat removal in channel regions, a coolant transport model, a row-wise coolant energy balance, and a surrogate-based optimization framework. This formulation enables efficient exploration of cooling geometries while retaining the dominant physical mechanisms governing temperature distribution and hydraulic resistance.

\subsection{Computational Domain and Discretization}

The computational domain is a square substrate of dimensions:
\begin{equation}
L_x = L_y = 0.12~\text{m}
\end{equation}

This defines the physical size of the chip package being analyzed.
The domain is discretized using a structured Cartesian grid:
\begin{equation}
N_x = N_y = 200
\end{equation}

The grid provides sufficient spatial resolution for the thermal analysis while maintaining computational efficiency.
The grid spacing can be defined through the dimension. 
\begin{equation}
\Delta x = \frac{L_x}{N_x} \qquad
\Delta y = \frac{L_y}{N_y}
\end{equation}

All field variables are evaluated on this grid, allowing an efficient finite-difference solution for repeated candidate geometries.

\subsection{Chip Power Model}

The system contains two GPUs and one CPU. Heat generation is spatially uniform within each chip footprint. Let $\Omega_{gpuL}$, $\Omega_{gpuR}$, and $\Omega_{cpu}$ denote these regions. The surface heat flux is defined as
\begin{equation}
q(x,y)=
\begin{cases}
\dfrac{P_{gpu}}{A_{gpu}}, & (x,y)\in \Omega_{gpuL}\cup\Omega_{gpuR}\\[6pt]
\dfrac{P_{cpu}}{A_{cpu}}, & (x,y)\in \Omega_{cpu}\\[6pt]
0, & \text{otherwise}
\end{cases}
\end{equation}

This distributes the total chip power uniformly over each chip area, ensuring consistent heat generation within the GPU and CPU regions.

\subsection{Channel Geometry Representation}

The cooling channels are represented by a binary indicator function.
\begin{equation}
\phi(x,y)=
\begin{cases}
1, & (x,y)\in \Omega_c\\
0, & (x,y)\notin \Omega_c
\end{cases}
\end{equation}

This function distinguishes channel regions from solid substrate regions, allowing different physical models to be applied in each.

The interdigitated channel layout is parameterized using three discrete design variables.
\begin{equation}
\mathbf{x} = (N, W, E)
\end{equation}
where $N$ is the number of channels, $W$ is the channel width, and $E$ is the expansion width over chip regions. These parameters fully define the channel layout used in the simulations. To improve thermal performance, channel placement is biased toward GPU regions. Instead of uniform spacing, candidate channel locations are selected from a combined set including GPU edges, GPU centers, the CPU region, and uniformly spaced positions across the domain. This ensures higher channel density near dominant heat sources while maintaining flow connectivity.

\begin{figure*}[htbp]
\centering
\includegraphics[width=0.75\linewidth]{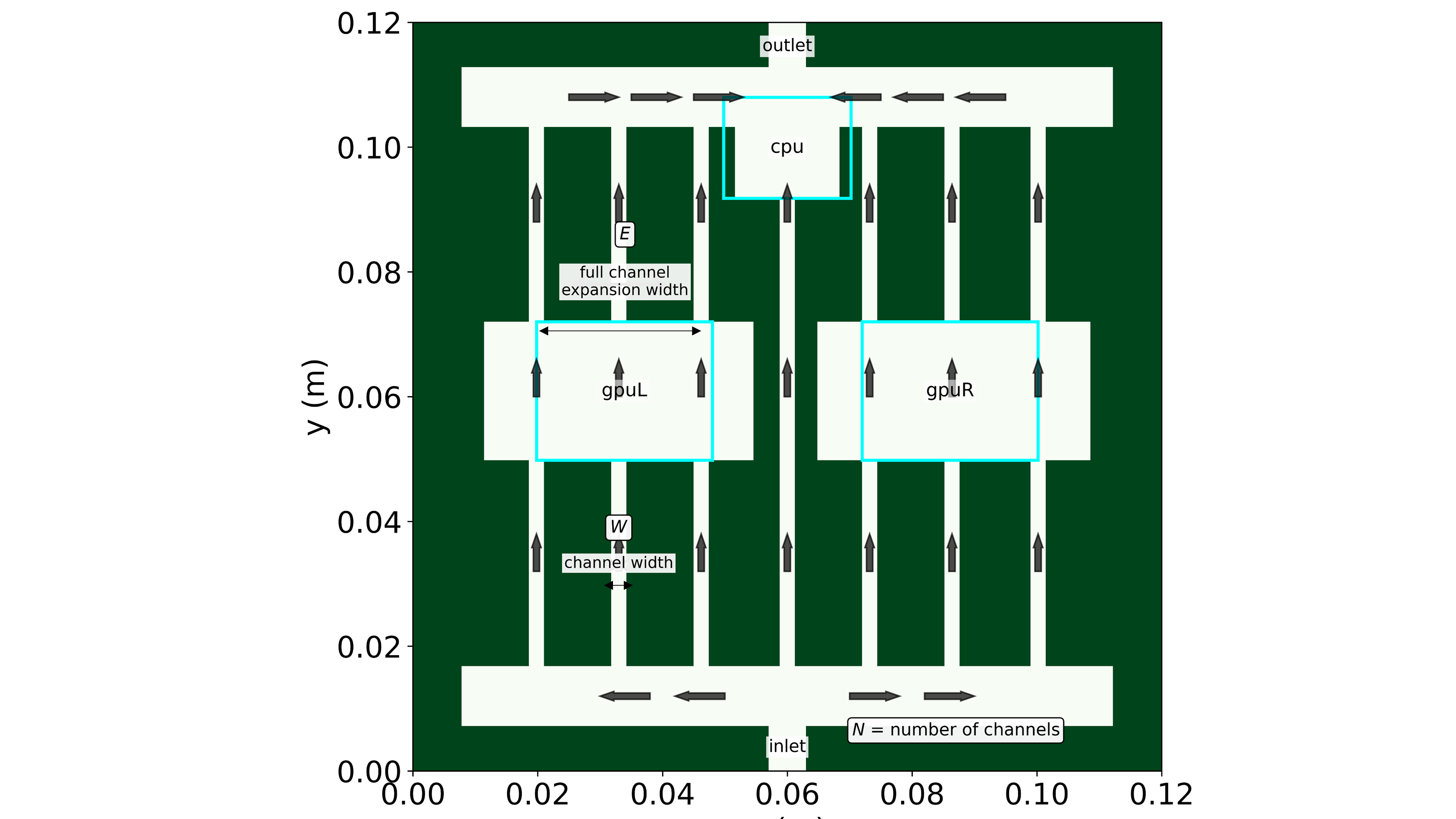}
\caption{Definition of the geometric design variables for the cooling manifold.}
\label{fig:parameters}
\end{figure*}

\subsection{Thermal Model}

The temperature distribution within the substrate is governed by a steady-state energy balance that couples conduction with convective heat removal. The governing equation is expressed as:
\begin{equation}
\nabla \cdot (k \nabla T)
-\frac{h_{eff}(x,y)}{t}\big(T-T_c\big)
+ q(x,y)=0
\end{equation}
where $T$ is the substrate temperature, $k$ is the thermal conductivity of the material, and $t$ represents the substrate thickness. The term $q(x,y)$ represents the spatial heat generation distribution derived from the chip power dissipation.

To account for the distinct cooling characteristics of the manifold and the surrounding substrate, the effective heat transfer coefficient is defined as:
\begin{equation}
h_{eff}(x,y)=h_{floor}+h_{chan}(x,y)
\end{equation}
where
$h_{floor}=0.05\,h_{ref}$ represents a weak background cooling level applied uniformly across the domain to account for secondary heat dissipation paths and improve numerical stability.

Enhanced cooling is restricted to the channel regions through
\begin{equation}
h_{chan}(x,y)=0.95 h_{ref} \phi(x,y)
\end{equation}
where $\phi(x,y)$ is the binary channel indicator function.
The binary indicator function restricts enhanced convective cooling to the channel regions while maintaining weak background heat removal throughout the substrate.

\begin{figure}[h]
\centering
\includegraphics[width=\linewidth]{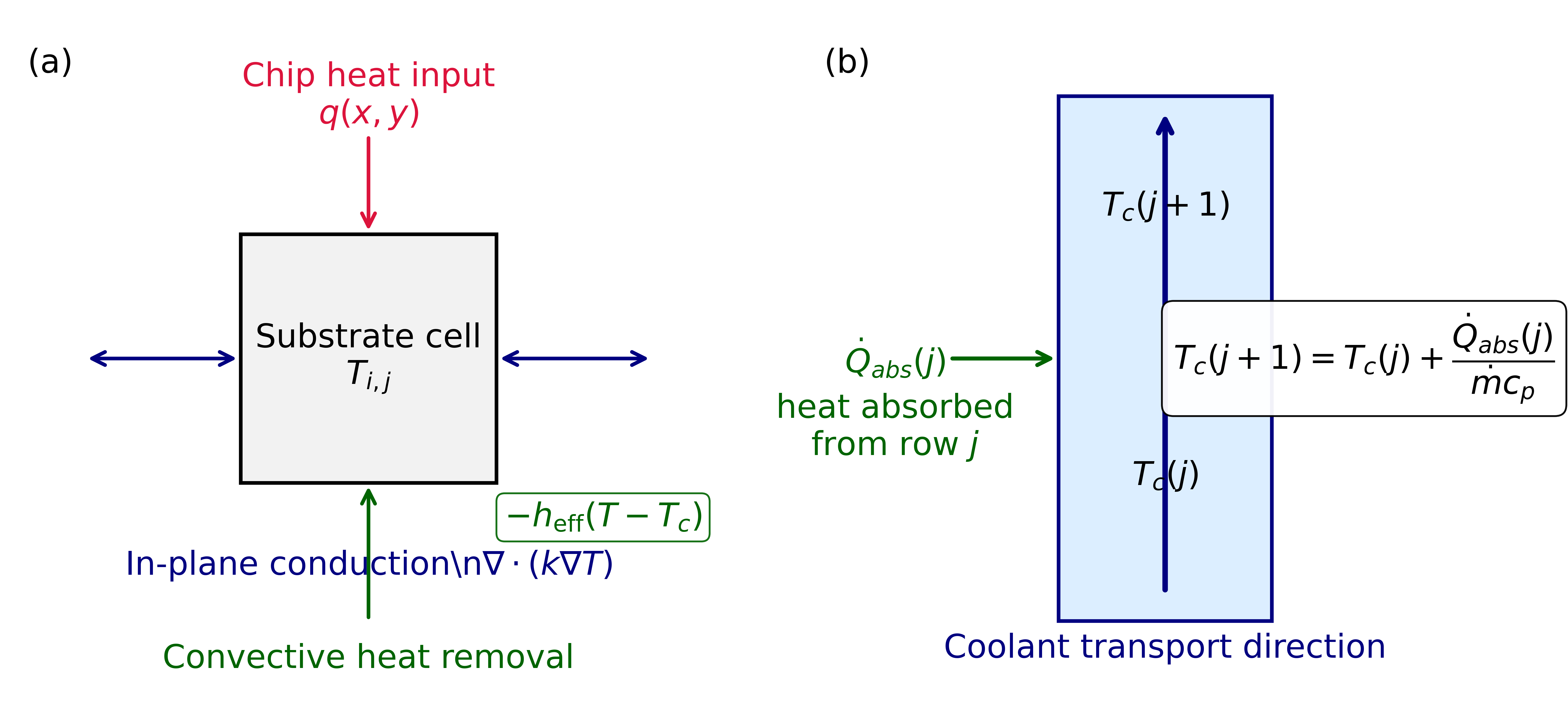}
\caption{(a) Schematic representation of the reduced-order thermal framework. (b) The substrate domain is modeled using two-dimensional heat conduction coupled with localized convective heat removal, while coolant heating is modeled using a row-wise energy transport approximation along the flow direction.}
\label{fig:coupling_framework}
\end{figure}

To improve computational efficiency while preserving the dominant thermal behavior of the system, the present framework couples a two-dimensional substrate conduction model with a reduced-order coolant transport approximation. Heat conduction within the substrate is modeled over the full computational domain, while coolant heating is approximated using a row-wise energy balance along the primary flow direction. Figure~\ref{fig:coupling_framework} illustrates the interaction between substrate heat conduction and coolant temperature evolution within the proposed framework.

\subsection{Coolant Energy Model}

As the fluid flows through the manifold, it gains heat. To account for this, we use a simple one-dimensional energy balance along the main flow direction. This helps us track how the coolant warms up as it moves downstream, which is important for finding possible hotspot near the outlet. The calculation starts at the inlet, where the coolant temperature in the first row of cells, $T_c(1)$ is set equal to the inlet temperature, $T_{in}$.
\begin{equation}
T_c(1)=T_{in}
\end{equation}

As the fluid moves through the channel network, the temperature at each successive row $j+1$ is updated based on the energy absorbed from the substrate in the preceding row $j$:
\begin{equation}
T_c(j+1)
=
T_c(j)
+
\frac{\dot{Q}_{abs}(j)}{\dot{m}c_p}
\end{equation}

This row-wise marching approach provides a computationally efficient approximation of the fluid temperature field, effectively coupling the global mass transport with local heat transfer.

The total heat absorbed at a specific row, $\dot{Q}_{abs}(j)$, is determined by summing the convective heat transfer from all channel-occupied grid cells $(i, j)$ across the $x$-direction:
\begin{equation}
\dot{Q}_{abs}(j)
=
\sum_{i=1}^{N_x}
h_{chan,i,j}\,
\max(T_{i,j}-T_c(j),0)\,
\Delta x \Delta y
\end{equation}
where $h_{chan,i,j}$ is the local convective heat transfer coefficient, $T_{i,j}$ is the substrate temperature at the corresponding grid node, and $\Delta x \Delta y$ represents the surface area of a single control volume. The use of a $\max$ function ensures that heat transfer is physically constrained to flow from the substrate to the coolant, preventing non-physical "cooling" of the substrate by a warmer fluid. This coupled energy balance ensures that the thermal model accounts for the reduced cooling capacity of the fluid as its temperature rises toward the outlet.

\subsection{Performance Metrics}

To assess the effectiveness of each candidate cooling layout, a set of metrics is calculated over the functional regions of the package. The thermal performance is primarily evaluated within the combined footprint of the GPUs and CPU, denoted as $\Omega_{chip}$. The peak thermal load is quantified by the maximum chip temperature, $T_{max}$, which is defined as the highest temperature found at any coordinate $(x,y)$ within the device regions:
\begin{equation}
T_{max}=\max_{(x,y)\in \Omega_{chip}} T(x,y)
\end{equation}

\begin{figure*}[h]
\centering
\includegraphics[width=0.75\linewidth]{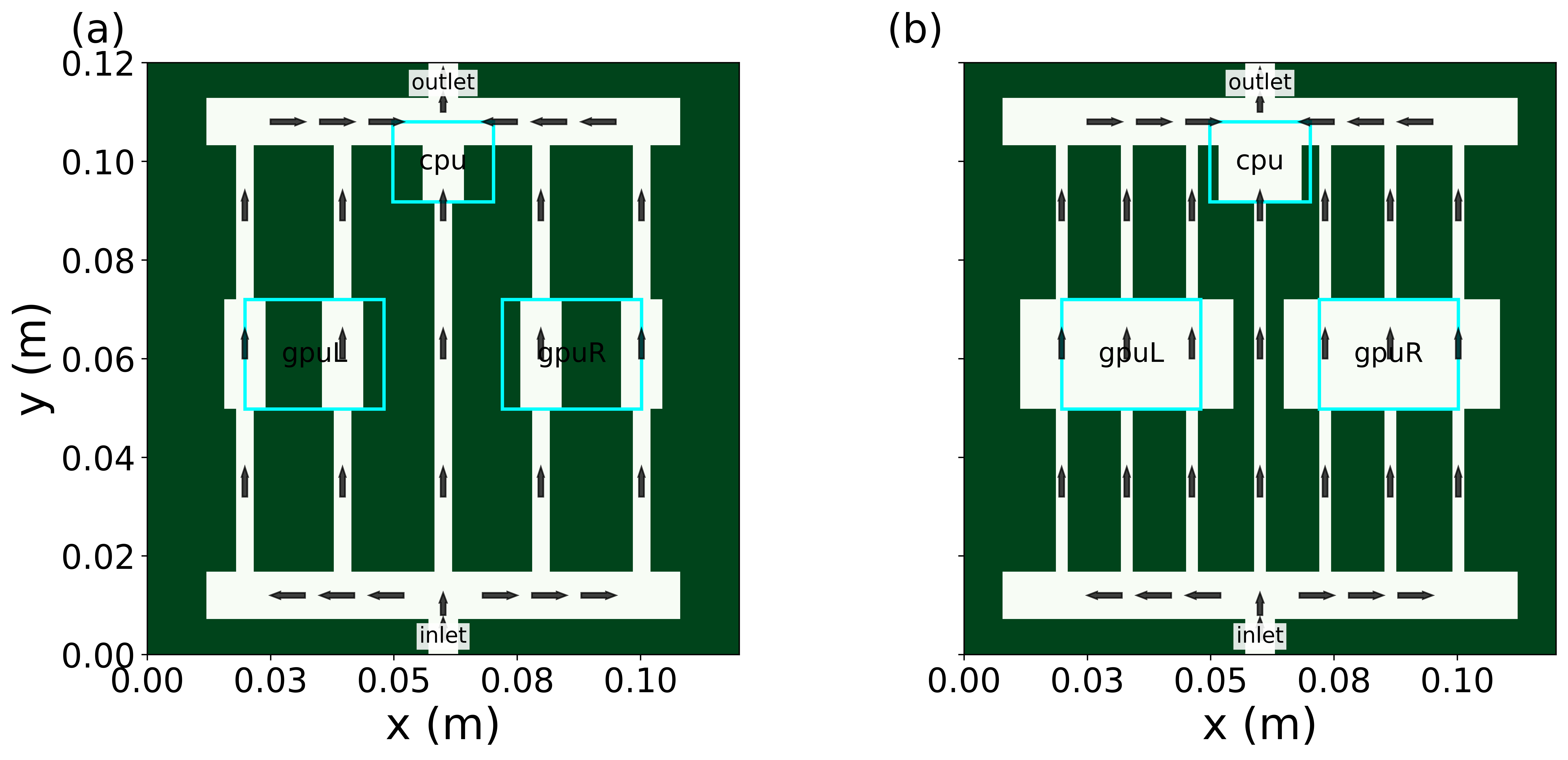}
\caption{Comparison between the baseline manifold-fed configuration (a) and the optimized cooling-channel layout; (b) The optimized design increases cooling coverage over the GPU and CPU regions through localized channel expansion while maintaining a symmetric coolant distribution throughout the package.}
\label{fig:topology}
\end{figure*}

While $T_{max}$ is essential for assessing reliability and preventing local device failure, the overall cooling efficiency is captured by the average chip temperature, $T_{avg}$. In our framework, this is computed as the arithmetic mean of the temperatures at all grid nodes $(i,j)$ belonging to the chip footprints:
\begin{equation}
T_{avg}
=
\frac{1}{N_{chip}}
\sum_{(i,j)\in \Omega_{chip}} T_{i,j}
\end{equation}
where $N_{chip}$ represents the total number of control volumes contained within $\Omega_{chip}$.

\subsection{Optimization}

The primary goal of this framework is to identify cooling topologies that mitigate both peak heat concentration and overall package thermal energy.
The optimization is formulated as a single-objective problem with the objective function:\begin{equation}\min_{\mathbf{x}} J(\mathbf{x})\end{equation}where the design vector $\mathbf{x} = (N, W, E)$ encompasses the geometric parameters of the manifold.
The objective function $J(\mathbf{x})$ is a multi-objective weighted combination of normalized thermal metrics:
\begin{equation}J(\mathbf{x}) = \omega_1 \frac{T_{avg}}{T_{avg}^{base}} + \omega_2 \frac{T_{max}}{T_{max}^{base}}\end{equation}

In this work, the weights are set to $\omega_1 = 0.3$ and $\omega_2 = 0.7$, prioritizing the reduction of the maximum junction temperature ($T_{max}$) to ensure device reliability, while still maintaining overall system efficiency via the average temperature ($T_{avg}$). Normalization by the baseline design values ($T^{base}$) ensures that both metrics are on a similar scalar magnitude, preventing one from dominating the optimization landscape.

Due to the non-linear coupling between heat conduction and fluid transport, direct optimization using the full finite-difference solver is computationally prohibitive. Consequently, a surrogate-based optimization approach is utilized to approximate the thermal response surface.A structured grid of $M=116$ candidate designs was sampled across the design space using a full-factorial approach within the variable bounds. For each sample point, the full physics-based solver was executed. The resulting temperature metrics were used to train quadratic surrogate models using a least-squares regression framework:\begin{equation}\hat{T}(\mathbf{x}) = \beta_0 + \sum_{i=1}^3 \beta_i x_i + \sum_{i=1}^3 \sum_{j=i}^3 \beta_{ij} x_i x_j\end{equation}

The accuracy of these surrogates was quantified using the Root Mean Square Error (RMSE). The obtained RMSE values for $\hat{T}_{max}$ and $\hat{T}_{avg}$ were 8.051$^\circ$C and 2.442$^\circ$C, respectively, indicating that the quadratic response surfaces sufficiently capture the design space trends.

The optimization is executed using the Gurobi Optimizer, which handles the mixed-integer nature of the design variables. The problem is formulated as a Mixed-Integer Quadratic Program (MIQP):
\begin{align}
\min_{\mathbf{x}} \quad & J(\mathbf{x}) \label{eq:objective} \\
\text{subject to} \quad & N = 2K + 1, \quad K \in \{2, 3, 4, 5\} \label{eq:constraints} \\
& 4 \le W \le 14 \\
& 12 \le E \le 28 \\
& E \ge W
\end{align}

To ensure architectural symmetry about the center of the 120 mm package, $N$ is constrained to be an odd integer by the relation $N = 2K + 1$. The continuous variables $W$ and $E$ represent the channel and expansion widths in millimeters. The linear inequality constraint $E \ge W$ is enforced to maintain physical realizability, ensuring the expansion region over the chip footprints is at least as wide as the feeder channels. The optimization utilizes the Gurobi non-convex solver settings to guarantee global optimality of the resulting quadratic surface.

\section{Results and Discussion}

The proposed thermal framework was applied to evaluate multiple manifold-fed cooling-channel configurations and identify an optimized design using the surrogate-based optimization framework described in Section 3. Each configuration was parameterized using the design variables $(N,W,E)$, representing the number of channels, channel width, and local expansion width over chip regions, respectively.

\subsection{Baseline and Optimized Performance}

The optimization process aimed to minimize the normalized objective function $J$, defined as a weighted combination of maximum and average chip temperatures. The baseline configuration corresponded to a symmetric manifold-fed channel layout with moderate channel width and limited cooling coverage over the chip regions.

The baseline configuration yielded the following thermal performance:

\begin{equation}
T_{max}^{base} = 220.36^\circ\text{C}
\quad
T_{avg}^{base} = 102.14^\circ\text{C}
\end{equation}

Using the surrogate-based optimization framework, the best-performing configuration identified by the Gurobi optimizer was:

\begin{equation}
(N,W,E) = (7,4,28)
\end{equation}

The validated thermal performance of the optimized configuration was:

\begin{equation}
T_{max} = 79.91^\circ\text{C}
\quad
T_{avg} = 66.27^\circ\text{C}
\end{equation}

This corresponds to temperature reductions of:

\begin{equation}
\Delta T_{max} = 140.45^\circ\text{C}
\quad
\Delta T_{avg} = 35.87^\circ\text{C}
\end{equation}

Compared to the baseline configuration, the optimized design significantly reduces hotspot intensity while maintaining more uniform cooling across the chip regions. Most importantly, the optimized maximum temperature falls below the approximate $95^\circ$C operating threshold commonly associated with high-performance GPU systems.

\subsection{Optimization Behavior}

The optimization framework revealed that thermal performance strongly depends on the balance between channel density, channel width, and cooling coverage over chip regions.
Increasing the number of channels improves spatial cooling coverage and reduces the average distance between heat-generating regions and cooling pathways. Similarly, increasing the expansion width over the chips increases the effective convective cooling area directly above the GPU and CPU regions, which significantly reduces hotspot formation.

During optimization, the solver consistently favored larger expansion widths over chip regions, indicating that cooling coverage near high-power components plays a dominant role in temperature reduction. At the same time, the optimizer avoided excessively large channel widths, suggesting that balanced flow distribution is more beneficial than simply maximizing channel area.
Unlike the earlier discrete optimization approach, the present framework treats the channel width and expansion width as continuous variables within prescribed bounds. This allows smoother exploration of the design space and improves optimization flexibility.

Direct optimization using the full thermal model is computationally expensive because each candidate design requires solving the steady-state temperature field across the computational domain. To improve efficiency, a surrogate-based optimization framework was adopted.

A set of candidate channel configurations was first simulated using the full thermal solver. The resulting temperature metrics were then used to construct quadratic surrogate models relating the design variables $(N,W,E)$ to the predicted values of $T_{max}$ and $T_{avg}$.

The surrogate optimization problem was subsequently solved using the Gurobi Optimizer. The resulting optimized design was then validated using the full thermal model. The close agreement between predicted and validated temperatures confirms that the surrogate model provides a reasonable approximation of the underlying thermal behavior.

The surrogate-based approach significantly reduces optimization cost while still preserving the main physical trends governing thermal performance.

\begin{figure}[h]
\centering
\includegraphics[width=0.7\linewidth]{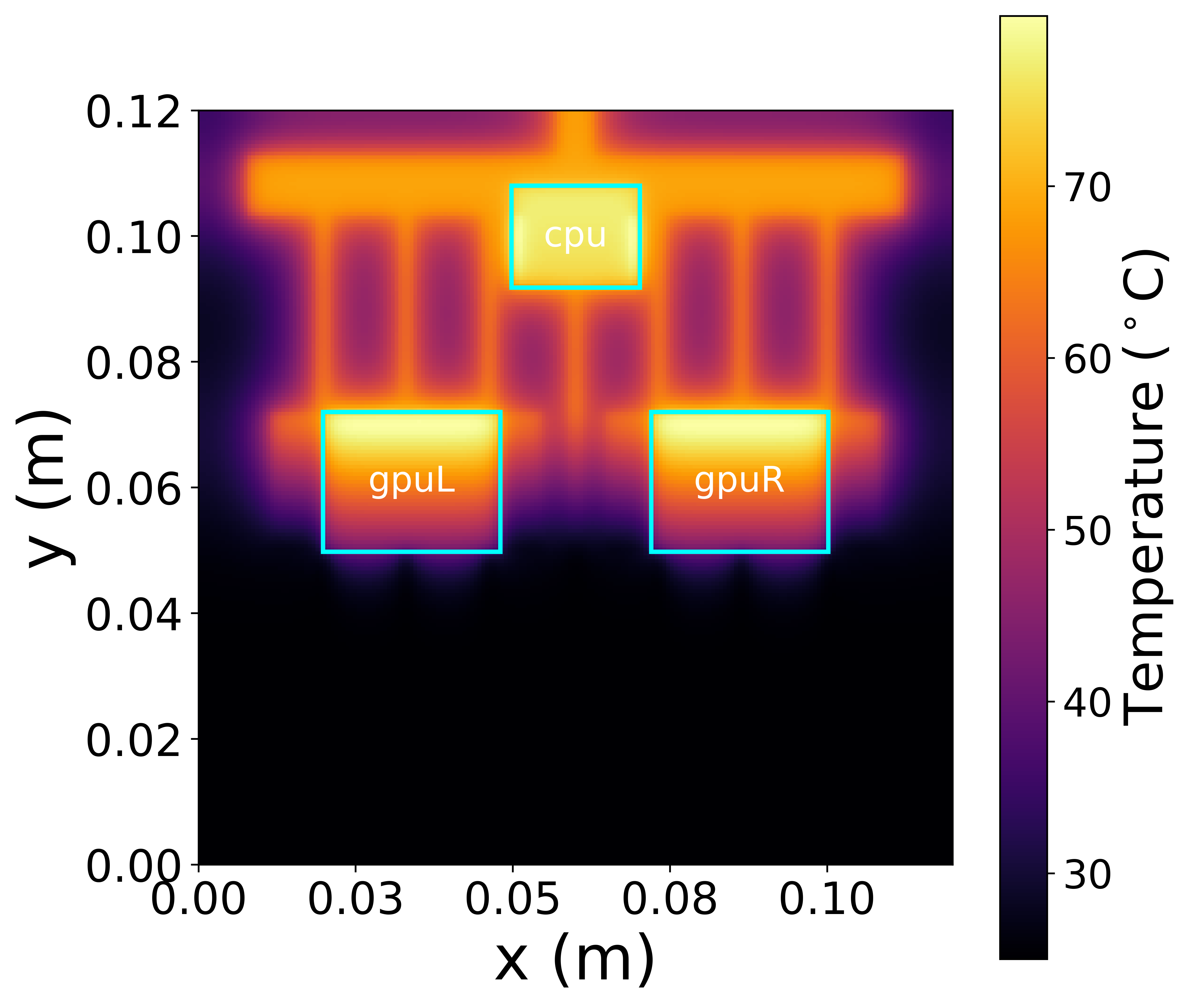}
\caption{Temperature contour field for the optimized cooling-channel configuration. The optimized design substantially reduces hotspot intensity while maintaining a more uniform thermal distribution across the multi-chip package.}
\label{fig:optimized_temperature}
\end{figure}

\subsection{Optimized Temperature Distribution}

The temperature field corresponding to the optimized cooling-channel configuration is shown in Fig.~\ref{fig:optimized_temperature}. The optimized design produces a substantially more uniform thermal distribution across the computational domain while significantly reducing hotspot intensity near the GPU regions.

Compared to the baseline configuration, the optimized channel layout improves local heat extraction by increasing cooling coverage directly above the high-power chip regions. As a result, the maximum chip temperature is reduced to below the approximate operating limit commonly associated with high-performance GPU systems.

\begin{figure*}[t]
\centering
\includegraphics[width=0.75\linewidth]{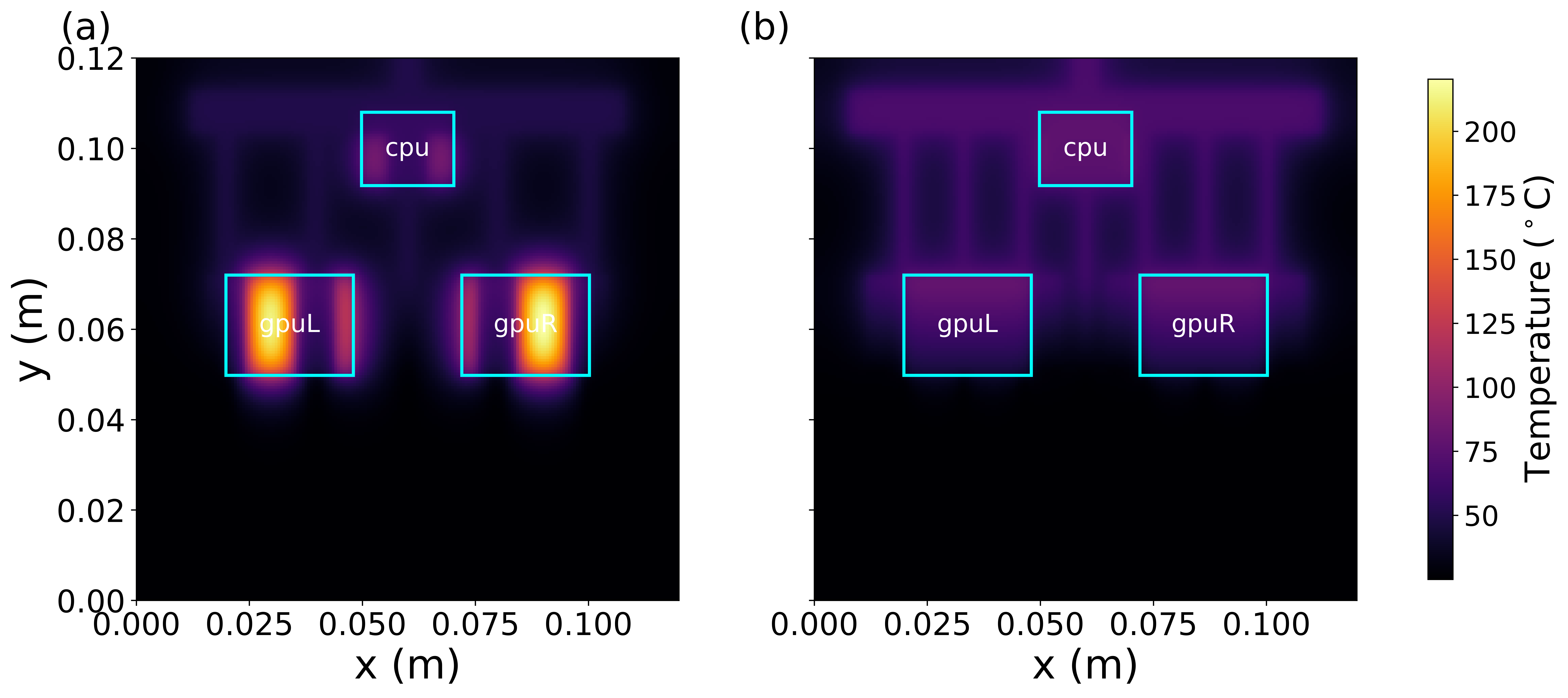}
\caption{Temperature contour comparison between the baseline configuration (a) and the optimized configuration (b). The optimized design significantly reduces hotspot intensity and produces a more uniform temperature field across the multi-chip package.}
\label{fig:temperature_comparison}
\end{figure*}

The temperature contours also demonstrate the effect of coolant heating along the flow direction. As the coolant absorbs heat from the substrate and chip regions, the upper portion of the domain exhibits slightly elevated temperatures near the outlet manifold. Nevertheless, the optimized configuration maintains substantially lower chip temperatures overall while preserving smooth and physically realistic thermal gradients throughout the substrate.

\subsection{Temperature Field Improvement}

Temperature contour plots demonstrate substantial improvement in thermal behavior for the optimized configuration. In the baseline design, strong hotspot formation occurs near the GPU regions due to concentrated heat generation and insufficient local cooling coverage.

In contrast, the optimized configuration produces a smoother thermal field with lower peak temperatures and reduced thermal gradients across the substrate. The expanded channel regions above the GPUs and CPU improve local heat extraction and reduce temperature accumulation near the hottest regions.

\begin{figure*}[t]
\centering
\includegraphics[width=\linewidth]{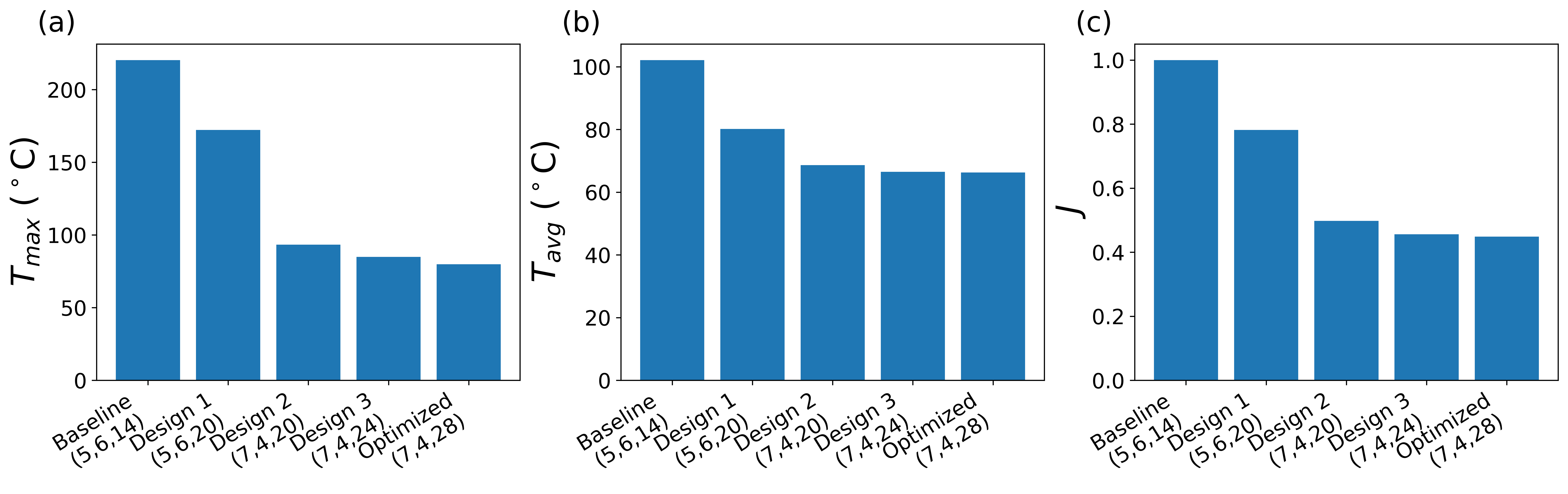}
\caption{Comparison of representative cooling-channel configurations showing (a) maximum chip temperature, (b) average chip temperature, and (c) normalized objective function value. The optimized design achieves the lowest objective value while simultaneously reducing both peak and average chip temperatures.}
\label{fig:representative_designs}
\end{figure*}

\subsection{Coolant Temperature Evolution}

The coolant temperature profile exhibits a monotonic increase along the flow direction as heat is absorbed from the substrate. Starting from the inlet temperature of approximately $25^\circ$C, the coolant temperature gradually rises as it passes through successive channel regions.

The optimized design exhibits a larger coolant temperature increase near the chip regions, indicating enhanced heat absorption and improved thermal extraction. This behavior confirms the importance of incorporating coolant temperature evolution within the thermal model. Neglecting coolant heating would underestimate downstream temperatures and overpredict cooling performance.

\begin{table}[htbp]
\centering
\caption{Thermal performance metrics for representative cooling-channel configurations.}
\label{tab:representative_designs}
\begin{tabular}{ccccccc}
\hline
Design & $N$ & $W$ & $E$ & $T_{max}$ ($^\circ$C) & $T_{avg}$ ($^\circ$C) & $J$ \\
\hline
Baseline & 5 & 6 & 14 & 220.36 & 102.14 & 1.000 \\
Design 1 & 5 & 6 & 20 & 172.28 & 80.18 & 0.782 \\
Design 2 & 7 & 4 & 20 & 93.40 & 68.63 & 0.498 \\
Design 3 & 7 & 4 & 24 & 84.90 & 66.50 & 0.456 \\
Optimized & 7 & 4 & 28 & 79.91 & 66.27 & 0.449 \\
\hline
\end{tabular}
\end{table}

Table~\ref{tab:representative_designs} quantitatively summarizes the thermal performance of the representative cooling-channel configurations considered during optimization.

\subsection{Representative Design Comparison}

To further evaluate the effect of the geometric design variables, several representative cooling-channel configurations were compared using the proposed thermal framework. The selected designs span different combinations of channel count, channel width, and local expansion width over the chip regions. For each configuration, the maximum chip temperature, average chip temperature, and normalized objective value were computed.

Figure~\ref{fig:representative_designs} compares the thermal performance metrics across the representative configurations. As the cooling coverage over the chip regions increases, both the peak and average chip temperatures decrease significantly. The optimized configuration achieves the lowest objective value while maintaining the lowest overall chip temperatures among the evaluated designs.

\subsection{Overall Performance Trends}

The results demonstrate that optimized cooling performance emerges from a balance between channel density, channel width, and cooling coverage over the chips. Increasing channel density improves spatial cooling distribution, while local expansion over chip regions enhances heat removal from the highest-power components.

The optimized design consistently favors increased cooling coverage near the GPUs, confirming that hotspot mitigation is strongly influenced by local convective area. At the same time, the optimization avoids excessively large channel widths, indicating that balanced coolant distribution remains important for maintaining thermal uniformity.

Overall, the surrogate-based optimization framework successfully identifies cooling-channel configurations capable of substantially reducing both peak and average chip temperatures under realistic operating conditions. These results demonstrate the effectiveness of combining reduced-order thermal modeling with mathematical optimization for topology-aware cooling design in advanced multi-chip electronic systems.
As shown in Fig. 7, the coolant temperature undergoes a significant gradient over the GPU regions ($0.05 \text{m} \le y \le 0.075 \text{m}$), where the localized expansion $E$ facilitates maximum heat uptake. The convergence of the slope toward the outlet manifold indicates that the majority of the $2700~\text{W}$ load is effectively absorbed within these optimized expansion zones.

\begin{figure}[h]
\centering
\includegraphics[width=0.95\linewidth]{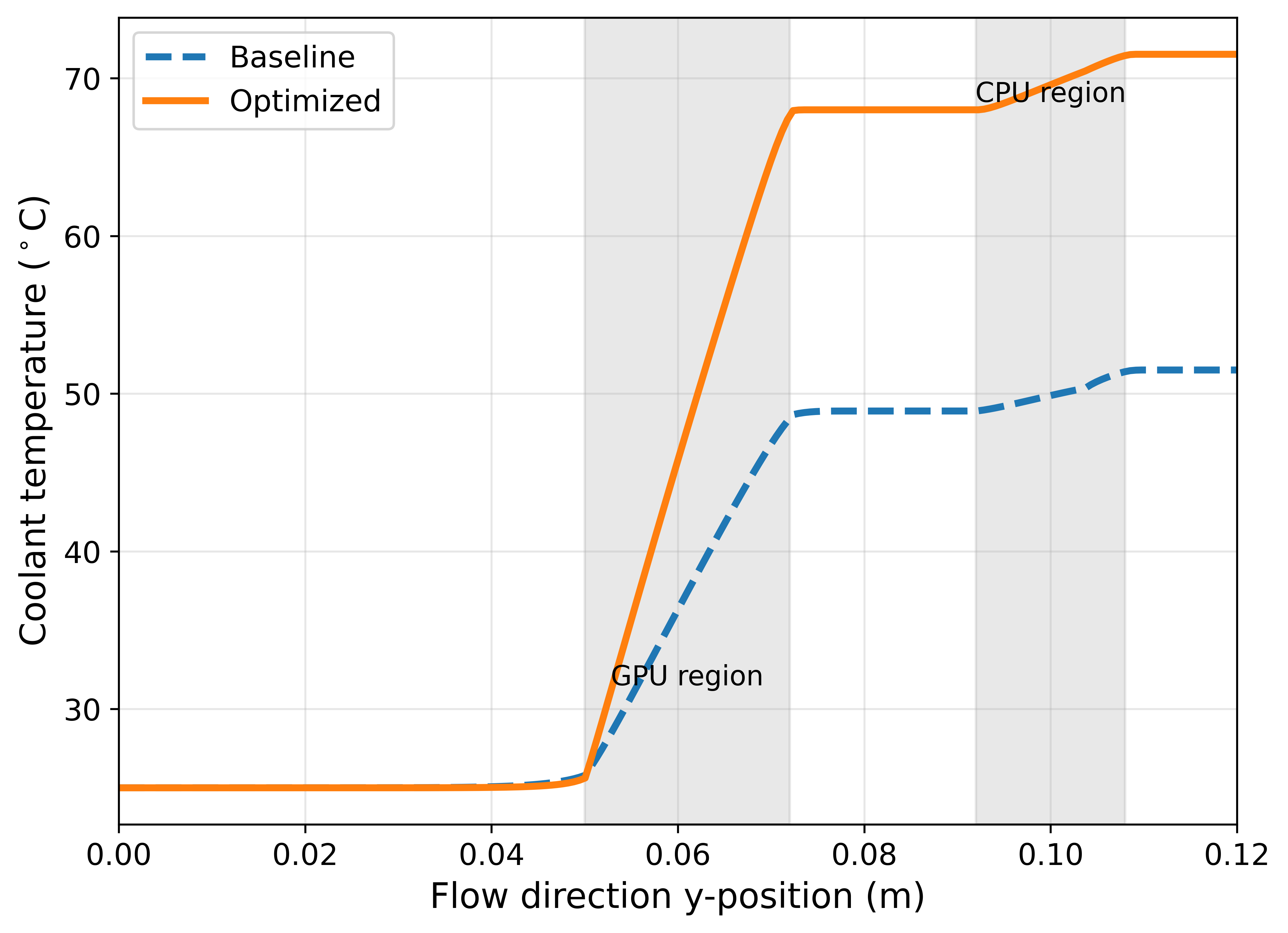}
\caption{Coolant temperature evolution along the flow direction for the optimized configuration. The increase in coolant temperature reflects heat absorption from the substrate and chip regions.}
\label{fig:coolant_profile}
\end{figure}

\section{Conclusion}

This study presented a computational framework for optimizing manifold-fed microchannel cooling designs for high-power multi-chip electronic systems. A reduced-order thermal model combining steady-state heat conduction, convective cooling, and a row-wise coolant energy balance was developed to evaluate temperature distributions within explicitly defined cooling-channel networks.
To improve cooling performance, the channel layout was parameterized using the number of channels, channel width, and local expansion width over chip regions. A surrogate-based optimization framework was then employed to efficiently approximate the relationship between channel geometry and thermal performance metrics. The resulting optimization problem was solved using a mathematical programming solver and validated using the full thermal model.
The optimized configuration achieved substantial reductions in both maximum and average chip temperatures relative to the baseline design. Most importantly, the optimized maximum temperature was reduced to below the approximate operating limit of modern high-performance GPU systems, demonstrating the effectiveness of increased cooling coverage near high-power regions.

Overall, the results show that combining thermal modeling with optimization driven channel design provides an effective and computationally efficient approach for  thermal management in advanced multi-chip packages. Future work may include higher-fidelity CFD validation, transient thermal analysis, and extension of the framework to larger multi-chip and data-center-scale cooling systems.

\section    *{ACKNOWLEDGMENTS}

This research is partially supported from the University of Michigan-Dearborn Office of Research through Research Initiation \& Development (RID) Grant.

\bibliographystyle{asmeconf}  
\bibliography{asmeconf-sample}

@article{liu2026generative,
  title={Generative Design for Direct-to-Chip Liquid Cooling for Data Centers},
  author={Liu, Zheng},
  journal={arXiv preprint arXiv:2604.10941},
  year={2026}
}

@misc{fibermallDeepDive,
	author = {FiberMall},
	title = {{D}eep {D}ive into {N}{V}{I}{D}{I}{A} {G}{B}200 {L}iquid {C}ooling {P}late {D}esign: {A}dvanced {L}iquid {C}ooling for {A}{I} {C}hips},
	howpublished = {\url{https://www.fibermall.com/blog/nvidia-gb200-liquid-cooling-plate.htm}},
	year = {2025},
	note = {[Accessed 22-03-2026]},
}

@article{tuckerman2005high,
  title={High-performance heat sinking for VLSI},
  author={Tuckerman, David B and Pease, Roger Fabian W},
  journal={IEEE Electron device letters},
  volume={2},
  number={5},
  pages={126--129},
  year={2005},
  publisher={IEEE}
}

@article{sobhan2001comparative,
  title={A comparative analysis of studies on heat transfer and fluid flow in microchannels},
  author={Sobhan, Choondal B and Garimella, Suresh V},
  journal={Microscale Thermophysical Engineering},
  volume={5},
  number={4},
  pages={293--311},
  year={2001},
  publisher={Taylor \& Francis}
}

@article{kandlikar2005high,
  title={High flux heat removal with microchannels—a roadmap of challenges and opportunities},
  author={Kandlikar, Satish G},
  journal={Heat Transfer Engineering},
  volume={26},
  number={8},
  pages={5--14},
  year={2005},
  publisher={Taylor \& Francis}
}

@article{colgan2007practical,
  title={A practical implementation of silicon microchannel coolers for high power chips},
  author={Colgan, Evan G and Furman, Bruce and Gaynes, Michael and Graham, Willian S and LaBianca, Nancy C and Magerlein, John H and Polastre, Robert J and Rothwell, Mary Beth and Bezama, RJ and Choudhary, Rehan and others},
  journal={IEEE transactions on components and packaging technologies},
  volume={30},
  number={2},
  pages={218--225},
  year={2007},
  publisher={IEEE}
}

@article{mudawar2002assessment,
  title={Assessment of high-heat-flux thermal management schemes},
  author={Mudawar, Issam},
  journal={IEEE transactions on components and packaging technologies},
  volume={24},
  number={2},
  pages={122--141},
  year={2002},
  publisher={Ieee}
}

@inproceedings{liang2020high,
  title={High efficiency liquid cooling system of power electronic converter},
  author={Liang, Jinhua and Xu, Haiping and Yuan, Zengquan and Zhou, Peng},
  booktitle={2020 5th Asia Conference on Power and Electrical Engineering (ACPEE)},
  pages={1270--1275},
  year={2020},
  organization={IEEE}
}

@article{escher2009efficiency,
  title     = {Efficiency of optimized bifurcating tree-like and parallel microchannel networks in the cooling of electronics},
  author    = {Escher, W. and Michel, B. and Poulikakos, D.},
  journal   = {International Journal of Heat and Mass Transfer},
  volume    = {52},
  number    = {5-6},
  pages     = {1421--1432},
  year      = {2009},
  publisher = {Elsevier}
}

@article{zhao2002analysis,
  title={Analysis of microchannel heat sinks for electronics cooling},
  author={Zhao, CY and Lu, TJ},
  journal={International Journal of Heat and Mass Transfer},
  volume={45},
  number={24},
  pages={4857--4869},
  year={2002},
  publisher={Elsevier}
}

@inproceedings{van2020embedded,
  title={Embedded microchannel cooling for high power-density GaN-on-Si power integrated circuits},
  author={Van Erp, Remco and Kampitsis, Georgios and Nela, Luca and Ardebili, Reza Soleimanzadeh and Matioli, Elison},
  booktitle={2020 19th IEEE Intersociety Conference on Thermal and Thermomechanical Phenomena in Electronic Systems (ITherm)},
  pages={53--59},
  year={2020},
  organization={IEEE}
}

@article{peles2005forced,
  title={Forced convective heat transfer across a pin fin micro heat sink},
  author={Peles, Yoav and Ko{\c{s}}ar, Ali and Mishra, Chandan and Kuo, Chih-Jung and Schneider, Brandon},
  journal={International Journal of Heat and Mass Transfer},
  volume={48},
  number={17},
  pages={3615--3627},
  year={2005},
  publisher={Elsevier}
}

@article{thome2004boiling,
  title={Boiling in microchannels: a review of experiment and theory},
  author={Thome, John R},
  journal={International Journal of Heat and Fluid Flow},
  volume={25},
  number={2},
  pages={128--139},
  year={2004},
  publisher={Elsevier}
}

@article{natarajan2009thermal,
  title={Thermal and power challenges in high performance computing systems},
  author={Natarajan, Venkat and Deshpande, Anand and Solanki, Sudarshan and Chandrasekhar, Arun},
  journal={Japanese Journal of Applied Physics},
  volume={48},
  number={5S2},
  pages={05EA01},
  year={2009}
}

@article{kong2023additively,
  title={An additively manufactured manifold-microchannel heat sink for high-heat flux cooling},
  author={Kong, Daeyoung and Jung, Euibeen and Kim, Yunseo and Manepalli, Vivek Vardhan and Rah, Kyupaeck Jeff and Kim, Han Sang and Hong, Yongtaek and Choi, Hyoung Gil and Agonafer, Damena and Lee, Hyoungsoon},
  journal={International Journal of Mechanical Sciences},
  volume={248},
  pages={108228},
  year={2023},
  publisher={Elsevier}
}

@article{wei2003optimization,
  title={Optimization study of stacked micro-channel heat sinks for micro-electronic cooling},
  author={Wei, Xiaojin and Joshi, Yogendra},
  journal={IEEE transactions on components and packaging technologies},
  volume={26},
  number={1},
  pages={55--61},
  year={2003},
  publisher={IEEE}
}

@inproceedings{harpole1991micro,
  title={Micro-channel heat exchanger optimization},
  author={Harpole, George M and Eninger, James E and others},
  booktitle={Proceeding of the 7th IEEE SEMI-THERM Symposium},
  pages={59--63},
  year={1991},
  organization={IEEE}
}

@article{song2022case,
  title={Case-embedded cooling for high heat flux microwave multi-chip array},
  author={Song, Yunqian and Fu, Rong and Chen, Chuan and Wang, Qidong and Su, Meiying and Hou, Fengze and Zhang, Xiaobin and Li, Jun and Cao, Liqiang},
  journal={Applied Thermal Engineering},
  volume={214},
  pages={118852},
  year={2022},
  publisher={Elsevier}
}

@article{zhu2024machine,
  title={Machine learning aided design and optimization of thermal metamaterials},
  author={Zhu, Changliang and Bamidele, Emmanuel Anuoluwa and Shen, Xiangying and Zhu, Guimei and Li, Baowen},
  journal={Chemical Reviews},
  volume={124},
  number={7},
  pages={4258--4331},
  year={2024},
  publisher={ACS Publications}
}

@article{chen2025additive,
  title={Additive manufacturing of vapor chambers},
  author={Chen, Kuan-Lin and Hsu, Shao-Chi and Kang, Shung-Wen},
  journal={Materials},
  volume={18},
  number={5},
  pages={979},
  year={2025},
  publisher={MDPI}
}

\end{document}